\documentclass[a4paper,
               ]{jacow}
\usepackage{pdfpages,multirow,ragged2e} %
\makeatletter%
	\ifboolexpr{bool{xetex}}
	 {\renewcommand{\Gin@extensions}{.pdf,%
	                    .png,.jpg,.bmp,.pict,.tif,.psd,.mac,.sga,.tga,.gif,%
	                    .eps,.ps,%
	                    }}{}
\makeatother

\ifboolexpr{bool{xetex} or bool{luatex}} 
 {}                                      
 {\usepackage[utf8]{inputenc}}           

\usepackage[USenglish]{babel}

\ifboolexpr{bool{jacowbiblatex}}%
 {%
  \addbibresource{jacow-test.bib}
  \addbibresource{biblatex-examples.bib}
 }{}
\begin{document}

\title{Development of an achromatic spectrometer \\ for a laser-wakefield-accelerator experiment}

\author{F.~Pe{\~n}a\textsuperscript{1}\thanks{felipe.pena@fys.uio.no}, E. Adli, P. Drobniak, D. Kalvik, K. N. Sjobak, C.~A.~Lindstr{\o}m,
    \\ Department of Physics, University of Oslo, Oslo, Norway \\ 
		\textsuperscript{1}also at Ludwig-Maximilians-Universit{\"a}t M{\"u}nchen, Munich, Germany}
	
\maketitle

\begin{abstract}
    The large gradients of plasma-wakefield accelerators promise to shorten accelerators and reduce their financial
    and environmental costs. For such accelerators, a key challenge is the transport of beams with high divergence
    and energy spread. Achromatic optics is a potential solution that would allow staging of plasma accelerators
    without beam-quality degradation. For this, a nonlinear plasma lens is being developed within the SPARTA
    project. As a first application of this lens, we aim to implement an achromatic spectrometer for electron
    bunches produced by a laser-wakefield accelerator. This will greatly improve the resolution across the typically one to tens of percent energy spread bunches and therefore help diagnosis and optimization of the plasma interaction. We report on progress in designing such an experiment.
\end{abstract}

\section{Introduction}

Plasma accelerators show a large potential for use in accelerator facilities thanks to their large accelerating gradients of up to 100~GV/m.
In its essence, a plasma wakefield accelerator transfers the energy of a driving bunch or laser pulse to a trailing bunch through the fields driven in the plasma.
For applications requiring high energies, many stages with a new driver at each stage will be required to accelerate a bunch.

The optics connecting the plasma-accelerator stages to each other are challenging~\cite{Lindstrom_PRAB_2021}.
The plasma provides strong focusing such that at the exit of the stage the beam is highly diverging.
Focusing the beam back into the next stage can introduce strong chromatic aberrations to the detriment of the beam emittance.
To ensure the compactness that arises from the high accelerating gradients of plasma-wakefield acceleration, the optics should be compact, requiring strong fields which can be beyond what conventional magnets can provide.

\begin{figure}[b]
   \centering
   \includegraphics*[width=1\columnwidth]{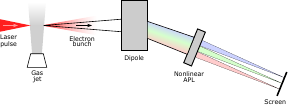}
   \caption{Experimental layout for an achromatically imaging spectrometer.
   The laser pulse enters the gas jet from the left and the laser-driven plasma-wakefield acceleration (LWFA) takes place.
   An electron bunch exits the gas at the right and drifts (200~mm) to a magnetic dipole (-0.85~T strong and 150~mm long), which bends its trajectory downward. The nonlinear plasma lens, 20~mm long and located 150~mm after the dipole and 400~mm before the screen, focuses the particles such that their position on the screen is independent of their initial angles.
   Red (blue) shows the fraction of the electron bunch with lower (higher) energy, and green the design energy; the schematic is not to scale.}
   \label{fig:experimental_setup}
\end{figure}

The SPARTA project~\cite{SPARTA,Lindstrom_IPAC_2025} is developing a nonlinear active plasma lens to provide viable staging optics, potentially solving one of the major remaining milestones in plasma-wakefield acceleration.
While proving a viable concept of staging is the final goal, the suitability of such a nonlinear active plasma lens can be demonstrated with a simpler application in plasma accelerators.

Laser-driven plasma-wakefield acceleration (LWFA) has been demonstrated across many different experiments and facilities with varying beam qualities, while they have usually charges in the range of pC--nC, divergence of <10~mrad, an energy spread from one to tens of percent, and a source size $\mathcal{O}$(1~\textmu m).
Measuring the energy spectrum of such bunches accurately is difficult due to the chromaticity of common optics.
While it can be done with multi-shot scans~\cite{Pena_PRR_2024}, such a method requires high stability, which is not usually the case.

In this article, we will show how such a nonlinear active plasma lens can be used in combination with a magnetic dipole to allow imaging an energy spectrum simultaneously throughout its energy range.
Such a setup would demonstrate the suitability of this device for the use in plasma-accelerated bunches, paving the way to its use in staging optics.

\begin{figure*}[h]
    \centering
    \includegraphics*[width=\textwidth]{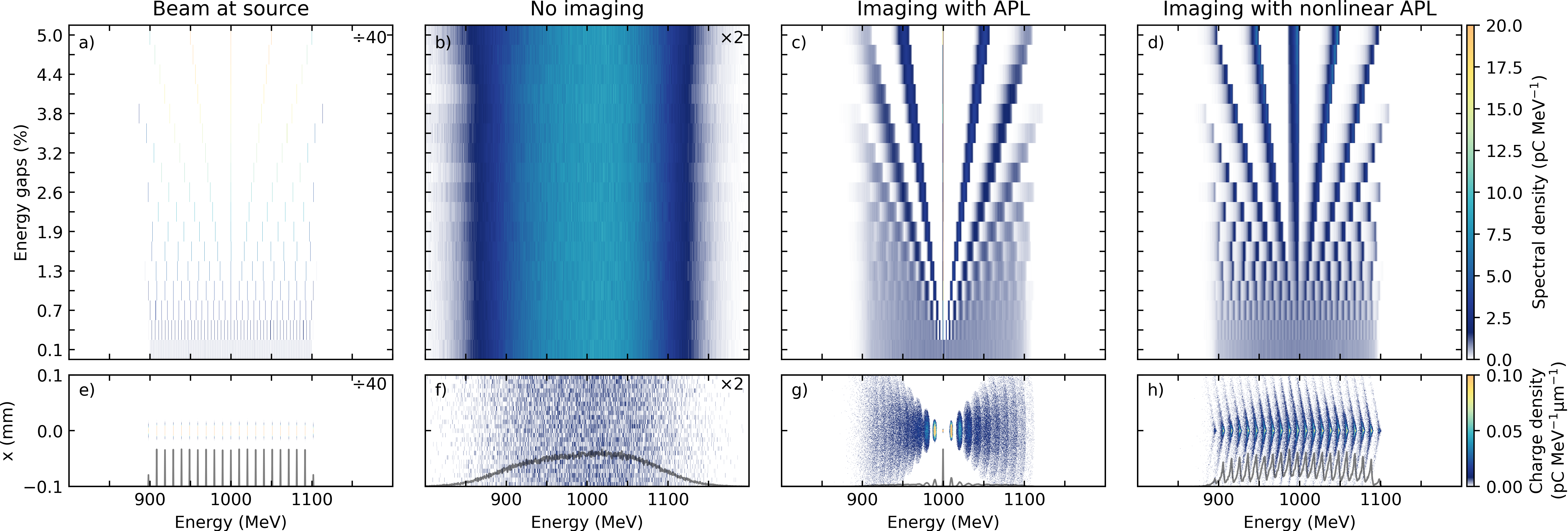}
    \caption{Simulated electron-bunch energy spectra at the source of the beamline (a), and at the spectrometer screen without imaging optics (b), imaging at 1~GeV with an APL (c), and imaging with the nonlinear APL.
    The electrons in the bunch bunch are initiated at the source with linearly-spaced discrete energies, and the relative separation is scanned.
    (e--f) show the images that would be measured on a screen at the end of the beamline to their corresponding beamline configuration, with the projection shown with the gray line.
    For these, bunches with the energy-gap of 1\% are selected.
    Note that the bins in the figures with `no imaging' and `beam at source' are scaled by a factor of 5 and 0.025, respectively, and that in (f) the bunch is not fully shown.}
    \label{fig:Waterfalls}
\end{figure*}

\section{Concept}

The plasma lens consists of a gas-filled sapphire capillary with electrodes at its ends~\cite{vanTilborg2015,Lindstrom2018}.
Applying a high voltage at the electrodes allows to initiate a discharge current, which will produce an azimuthal magnetic field and focus a passing charged particle bunch.
Adjusting the current allows changing the focusing strength of such an active plasma lens (APL).
Applying an external dipole magnetic field changes the radially uniform current via the Hall effect \cite{Drobniak_NuclInstr_2025, Drobniak_IPAC_2025}.
What would be a radially symmetric and linear field is now only linear on one axis, while in the other it is nonlinear (linearly increasing in gradient from one side to the other).
A beam dispersed on the same axis as the nonlinearity in the magnetic field will then experience a position-varying focusing force.
This nonlinearity can, in principle, be tailored via the external magnetic field to provide point-to-point imaging for the full energy range of the bunch simultaneously.
A sketch of such an achromatically imaging spectrometer, or in short \textit{achromatic spectrometer}, is shown in Fig.~\ref{fig:experimental_setup}.

\section{Comparison between different spectrometer setups}

In this section, we will compare different types of electron spectrometers using the ABEL~\cite{ABEL} simulation framework and the tracking code ImpactX~\cite{ImpactX}.
With a beamline as shown in Fig.~\ref{fig:experimental_setup}, the electron bunch at source has a normalized emittance $\epsilon_x = \epsilon_y = $ 2~mm~mrad, a waist beta function $\beta^*_x = \beta^*_y = $ 1~mm, a relative energy spread of 10\%~rms, a charge of 1~nC, and represented with \num{200000} particles.
To imprint visible features in the energy spectrum at the spectrometer screen, the energies of the electrons were rearranged to equally-spaced discrete values.
Scanning the spacing between these discrete values allows refining the spectrum features, e.g., from 5\% to 0.1\% as shown in Figs.~\ref{fig:Waterfalls}(a) and (e).

Often the main diagnostic in an LWFA experiment is an electron spectrometer without imaging optics~\cite{Foerster_PRX_2022}.
The consequence is that the position at the screen will no longer depend only on the electron energy, but also on the particle's initial angle, leading to fading -- or blurring -- of features in the spectrum.
For low-energy-spread bunches, this leads to a larger apparent energy spread on screen.
These two undesired effects are clearly visible in Figs.~\ref{fig:Waterfalls}(b) and (f), where no clear features in the spectrum are discernible and the apparent energy spread measured on the spectrometer screen is larger than at the source.

To overcome the limitations of a non-imaging setup, some experiments use a set of magnetic quadrupoles, which allow point-to-point imaging from the plasma source to the screen~\cite{Weingartner_PRAB_2011, Barber_PRL_2017, Pena_PRR_2024}.
The magnetic quadrupoles can be exchanged with a single APL, which is focusing in both transverse planes, to the same result.
As both types of imaging optics are chromatic, point-to-point imaging only works for a given \textit{imaging energy}.
Only close to this energy the spectrum is well imaged, and further away the divergent angles of the particles lead to blurring of the spectrum.
The blurring effect is exacerbated by the imaging optics compared to a setup without imaging.
This can be seen in Figs.~\ref{fig:Waterfalls}(c) and (g), where only the features in the energy spectrum around the imaged energy are well resolved throughout the scan.
Unless the energy spread is very small and the energy stability very high, the full spectrum cannot be measured precisely.
However, such a setup still allows measuring various beam parameters with different techniques, such as: emittance through a multi-shot imaging energy scan~\cite{Lindstrom_NatComm_2024} or a single-shot `butterfly' method~\cite{Weingartner_PRAB_2012}, and the beta function, the waist size and the beam divergence with multiple shots~\cite{Lindstrom_PRAB_2020}.

\section{Imaging with the nonlinear plasma lens compared to conventional imaging}

To ensure that particles of all energies are point-to-point imaged on the screen simultaneously, they have to be focused onto the screen with varying focusing strength.
The magnetic dipole disperses the bunch, i.e., the particles' vertical position is determined by their energy.
In such a location with an energy-dispersed beam, a plasma lens with nonlinear focusing matched to the local dispersion value and plane is well suited to provide the varying focusing strength required to correctly image all the energies present in the bunch onto the screen.
Typical values for the nonlinear plasma lens are 100--1000~T/m with a nonlinearity of 1\%~\cite{Drobniak_NuclInstr_2025}.
Figure~\ref{fig:Waterfalls}(h) shows the charge density that would be measured at a screen after such a setup, and (d) shows that the features in the spectrum can be seen down to features at the 0.5\% level.
While the APL can image smaller features close around its imaging energy, the nonlinear APL is able to resolve the full spectrum with similar resolution.

\section{Point-spread function}

\begin{figure}[!htb]
   \centering
   \includegraphics*[width=1\columnwidth]{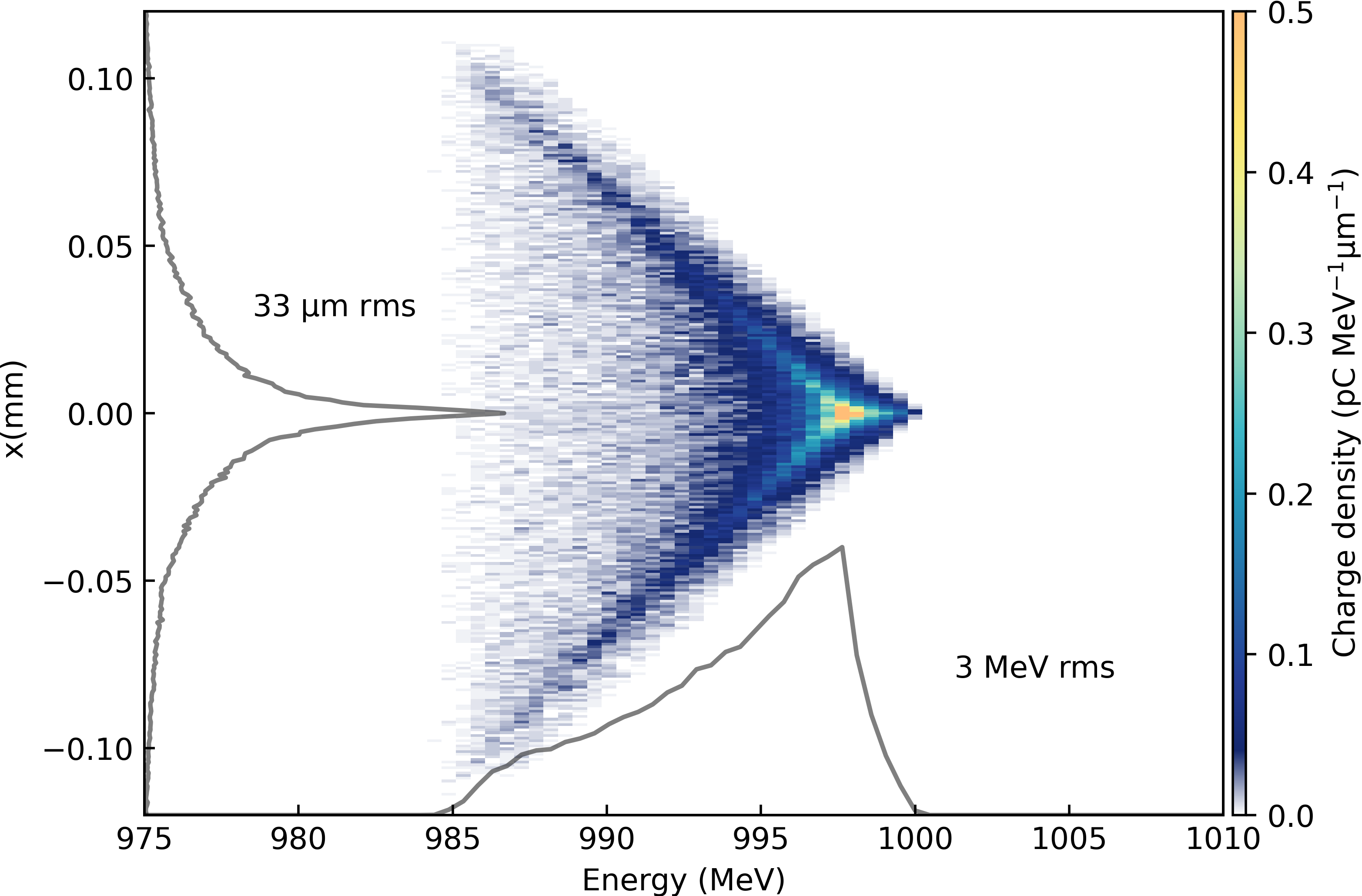}
   \caption{Shape of a beamlet on the screen, showing the PSF of the nonlinear-APL imaging system. The gray lines show the projections of the charge density.}
   \label{fig:image_PSF}
\end{figure}

As the focusing is symmetric, an APL yields round beamlets at the screen [see Fig.~\ref{fig:Waterfalls}(g)].
However, this is clearly different for the nonlinear APL, due to the strongly nonlinear fields.
To show this more clearly, we show in Fig.~\ref{fig:image_PSF} a zoomed-in image of one such beamlet.
This shape can be considered as the point-spread-function (PSF) of this idealized imaging system, i.e., a realistic bunch will be seen at the screen as a convolution with this shape.
Such a PSF limits the resolution of the achromatic-spectrometer setup.
According to Fig.~\ref{fig:image_PSF}, with the parameters used for the simulation, the rms width of 3~MeV shows that features down to the 0.3\% level should be visible, which is usually below the energy spread of bunches in LWFA and beam-driven plasma-wakefield acceleration.
In principle, the measurement can be de-convolved with the PSF to achieve higher resolution.

\section{Plans for experimental implementation at CALA}

To demonstrate the efficacy of such an imaging optic in a plasma-accelerator setup, an experiment is being designed at the Centre for Advanced Laser Applications (CALA)~\cite{CALA}, where currently no imaging optics are used.
This addition would allow measuring accurately the full energy spectrum of an LWFA bunch in a single shot.
Adjusting the focusing strength to image different object planes, it is possible to extract the transverse geometric emittance as in Ref.~\cite{Lindstrom_NatComm_2024}.
With the currently envisaged (tunable) design of the nonlinear APL, it will also be possible to switch off the external field and use the device as a normal APL, which would allow comparing the novel device with more established optics, and also measuring the emittance with a single shot~\cite{Weingartner_PRAB_2012}.

The LWFA-generated bunches at CALA currently have few-mrad divergence, $\sim$10\% energy spread, and up to 2~GeV energy.
Assuming the diameter of the nonlinear APL to be technically limited to 2~mm, care has to be taken to not collimate the beam with the capillary aperture.
In the results shown in Fig.~\ref{fig:Waterfalls}, a lens diameter of 1~mm was used, which lead to a charge loss of 30\%.
To fit such a beam fully inside this aperture, the lens has to be placed in a position where the beam dispersion is at most $\frac{1~\text{mm}}{10\%} = 10$~mm, consistent with the nonlinear plasma lenses currently being developed \cite{Drobniak_NuclInstr_2025}.

The experiment at CALA is scheduled to start in late 2026.

\section{Conclusion}

We have shown with simulations that a novel nonlinear APL developed in the SPARTA project could point-to-point image electron bunches achromatically across a $\pm$10\% energy spectrum.
While the point-spread-function of such imaging optics differs from common optics, based on the simulation parameters used in these results, it could be possible to measure features as small as at the 0.3\% level simultaneously across a broad energy spectrum.

\section{Acknowledgments}
This work was funded by the European Research Council (ERC Grant Agreement No. 101116161).
We acknowledge Sigma2 -- the National Infrastructure for High-Performance Computing and Data Storage in Norway for awarding this project access to the LUMI supercomputer, owned by the EuroHPC Joint Undertaking, hosted by CSC (Finland) and the LUMI consortium.

%
%
\ifboolexpr{bool{jacowbiblatex}}%
	{\printbibliography}%
	{%
	
	
} 

\end{document}